\documentclass[aps,prl,twocolumn,showpacs,superscriptaddress]{revtex4}
\usepackage{graphicx}
\begin{document}


\title{Phase separation and suppression of critical dynamics at 
quantum transitions of itinerant magnets: MnSi and (Sr$_{1-x}$Ca$_{x}$)RuO$_{3}$}
     \author{Y.~J.~Uemura}
     \affiliation{Department of Physics, Columbia University, New York, New York 10027, USA}
     \author{T.~Goko}
     \affiliation{Department of Physics, Tokyo University of Science, Noda, Chiba 278-8510, Japan}
     \affiliation{Department of Physics, Columbia University, New York, New York 10027, USA}
     \author{I.~M.~Gat-Malureanu}
     \affiliation{Department of Physics, Columbia University, New York, New York 10027, USA}
     \affiliation{Department of Science, SUNY Maritime College, 
Throggs Neck, New York, NY 10465, USA}
     \author{J.~P.~Carlo}
     \author{P.~L.~Russo}
     \author{A.~T.~Savici}
     \affiliation{Department of Physics, Columbia University, New York, New York 10027, USA}
     \author{A.~Aczel}
     \author{G.~J.~MacDougall}
     \author{J.~A.~Rodriguez}
     \author{G.~M.~Luke}
     \author{S.~R.~Dunsiger}
     \affiliation{Department of Physics and Astronomy, McMaster Univ., Hamilton, ON, Canada}
     \author{A.~McCollam}
     \affiliation{Department of Physics, University of Toronto, Toronto M5S 1A7, Ontario, Canada}
     \author{J.~Arai}
     \affiliation{Department of Physics, Tokyo University of Science, Noda, Chiba 278-8510, Japan} 
     \author{Ch.~Pfleiderer}
     \author{P.~B\"oni}
     \affiliation{Physik Department, Technische Universit\"at M\"unchen,
D-85748 Garching, Germany}
     \author{K.~Yoshimura}
     \affiliation{Department of Chemistry, Kyoto University, Kyoto 606-8502, Japan}
     \author{E.~Baggio-Saitovitch}
     \author{M.~B.~Fontes}
     \author{J.~Larrea J.} 
     \affiliation{Centro Brasilieiro de Pesquisas Fisicas, Rua Xavier Sigaud 150 Urca, CEP 22290-180
Rio de Janeiro, Brazil}
     \author{Y.~V.~Sushko}
     \affiliation{Department of Physics and Astronomy, University of Kentucky, Lexington, KY 40506-0055, USA}
     \author{J.~Sereni}
     \affiliation{Lab. Bajas Temperaturas, Centro Atomico Bariloche, Bariloche, Argentina}
\date{\today}

 

\maketitle

{\bf
Quantum phase transitions (QPTs) have been 
studied extensively in correlated electron systems.
Characterization of magnetism at QPTs has, however, been 
limited by the volume-integrated feature of neutron and magnetization 
measurements 
and by pressure uncertainties in NMR 
studies using powderized specimens.
Overcoming these limitations, we performed muon spin relaxation 
($\mu$SR) measurements which have a unique sensitivity to
volume fractions of magnetically ordered and paramagnetic regions,
and studied QPTs from itinerant heli/ferro magnet to paramagnet 
in MnSi (single-crystal; varying pressure) and 
(Sr$_{1-x}$Ca$_{x}$)RuO$_{3}$ (ceramic specimens; varying $x$).  
Our results provide the 
first clear evidence that both cases are associated with 
spontaneous phase separation and suppression of dynamic critical behavior,
revealed a slow but dynamic character of the
``partial order'' diffuse spin correlations in MnSi
above the critical pressure, and, combined with other known results
in heavy-fermion and cuprate systems, 
suggest a possibility that a majority of QPTs involve
first-order transitions and/or phase separation.\/}
\\

Advances of materials preparation and characterization techniques
have revealed complicated and sometimes unexpected phenomena near 
phase transitions and phase boundaries in correlated electron systems.
Fascinating examples can be found in pressure-tuned crossover from 
ferro (or heli) magnetic to paramagetic states in itinerant electron 
systems, such as MnSi \cite{pfleidererMnSiPRB,pfleidererMnSiNature} 
and ZrZn$_{2}$ \cite{pfleidererZrZn2Nature}, as well as
UGe$_{2}$ \cite{saxenaUGe2Nature} which is associated with appearance 
of a superconducting state near the disappearence of ferromagnetism.  
MnSi exhibits magnetic order with a spontaneous ordered moment
$m_{s}(T\rightarrow 0)$ = 0.4 Bohr magneton per Mn and 
a long period (180 \AA) helical modulation below $T_{c}$ = 29.5 K
at ambient pressure \cite{ishikawaMnSiPRB}.  This system has been 
extensively studied as a prototypical weak itinerant magnet existing 
near the boundary of disappearence of metallic ferromagnetism 
in the evolution from Fe, Ni, MnSi, to correlated 
paramagnet Pd, with decreasing degree of localized moment character
and decreasing ordered moment size $m_{s}$ relative to the effective 
paramagnetic moment $m_{p}^{eff}$ \cite{moriyaBook}.  
Magnetic order in MnSi can be suppressed by application of 
hydrostatic pressure $p$ above the critical pressure $p_{c}$ = 14.6 kbar \cite{pfleidererMnSiPRB}.
History dependence of magnetic susceptibility observed 
between $p^{*}$ = 12 kbar and $p_{c}$ suggests a first-order thermal phase 
transition at $T_{c}$ in a narrow pressure region before disappearance of 
magnetic order.
Recent neutron studies \cite{pfleidererMnSiNature} have revealed existence of 
``partial order'' spin correlations,
extending over a wide pressure region at $p > p_{c}$ at low temperatures
below $T_{o}$, whose diffusive intensity profile in reciprocal space is 
illustrated in Fig. 1(a).  Non-zero ordered moment size 
at $T \rightarrow 0$, measured by Si nuclear magentic resonance (NMR)
\cite{thessieuMnSiPhysica,yuMnSiPRL} up to $p = p_{c}$, indicates 
disappearance of magnetism in the first-order transition as a function of 
$p$.  NMR intensity from magnetically ordered regions was 
found \cite{yuMnSiPRL} to decrease with increasing $p$ at 
12 - 17 kbar.  Pressure inhomogeneity in powderized NMR 
specimens \cite{yuMnSiPRL}, however, prevented unambiguous determinations of 
(a) pressure regions with 
spatial heterogeneity, and (b) the 
relationship of this phenomenon to the partial order behavior.

With its unique sensitivity to slow spin fluctuations and to signals 
both from paramagnetic and ordered
volume fractions, muon spin relaxation ($\mu$SR) 
\cite{scottish,saviciLCOPRB} is a probe well suited to shed
new light on magnetic behaviors around QPT.  
Following earlier $\mu$SR studies of MnSi in ambient 
pressure \cite{hayanoMnSiPRL,kadonoMnSiPRB,gatMnSiPRL}, we explored the 
crossover region with $p$ = 10 - 16 kbar using a single crystal specimen in a 
standard piston type pressure cell with a backward muon beam of 
momentum 103 MeV/c from the M9 muon channel at TRIUMF.  
Typically $\sim$ 30 \%\ of the muons
stopped in the specimen, of 6 mm in diameter and 10 mm long, 
while 70 \%\ in the wall of the pressure cell, of 23 mm in outer diameter.
To assure temperature homogeneity, we used a gas-flow cryostat and two
independent thermometers located at the top/bottom of the pressure cell,
whose reading matched typically within +/- 0.1 K.

In single crystal specimens of ordered
magnetic systems, measurements in a weak transverse field (WTF) is the best 
way to obtain $\mu$SR signals from a paramagnetic volume fraction.
Figure 1(b) shows the precession signal and its envelope function observed 
in WTF = 100 G in $p$ = 15 kbar at T = 50 K and 2.5 K.
(Technical details of $\mu$SR data processing are described 
in the attached ``method'' section.)    The nearly identical 
signal at these two temperatures indicate that 100 \%\ of the volume is 
paramagnetic at T = 2.5 K in this pressure.
At $p$ = 9.6 kbar, as shown in Fig. 1(c), 
the envelope exhibits a clear reduction with decreasing temperature, 
caused by the missing signal from the magnetically ordered region of 
the specimen.  The pressure dependence of the signal at T = 2.5 K in 
Fig. 1(d) shows that the volume fraction of the magnetically ordered region 
increases with decreasing pressure.  
From earier $\mu$SR studies, it is known that 100\%\ of the volume
shows magnetic order below $T_{c}$ in ambient pressure 
with rather large internal fields (0.9 kG and 2.1 kG) 
at muon sites at low temperatures \cite{kadonoMnSiPRB}.  
Thus, the envelope signals for $p$ = 0 and 9.6
kbar in Fig. 1(d) represents muons stopped in the pressure cell.
  
By subtracting this background signal from the observed envelope, 
we can obtain the WTF signal from the 
paramgnetic region of the specimen, from which the volume
fraction $V_{M}$ of the region with static magnetic order was derived. 
Temperature and pressure dependence of $V_{M}$ is shown in Fig. 2(a) and (b).
In the pressure region between $p$ = 11.7 kbar ($\sim p^{*}$) and 13.9 kbar,
the static magnetic order remains in a partial volume fraction 
at $T \rightarrow 0$,
and the ordered region completely disappears at $p$ = 15 kbar, which is slightly 
above $p_{c}$ = 14.6 kbar.  
The frequency of spontaneous muon spin precession from the 
magnetically ordered volume, determined in separate zero-field $\mu$SR 
measurements,
remain finite below 13.9 kbar, as shown in Fig. 2(b).  These results  
indicate that the region between $p^{*}$ and $p_{c}$ is associated with 
phase separation between ordered and paramagnetic volumes, while there is no
volume with static magnetism above $p_{c}$.  Static magnetism with 
Mn spin component $>$ 0.004 $\mu_{B}$/Mn, either in ferromagnetic
or spin glass like correlations, would have produced static 
internal magnetic field $>$ 10 G (significantly larger than the nuclear dipolar fields $\sim 4$ G from Mn).
This situation should have lead to a distinguishable difference between T = 50 K and 2.5 K data in Fig. 1(b),
and thus can be ruled out by the present data. 

To study dynamic spin fluctuations, we also performed measurements of the
muon spin relaxation rate $1/T_{1}$ in a longitudianl field (LF) of 200 G.
As shown in Fig. 2(c), the relaxation rate at $p \sim$ 1 kbar exhibits
a divergent behavior, reproducing earlier results in ambient 
pressure \cite{hayanoMnSiPRL,kadonoMnSiPRB,gatMnSiPRL}. 
The critical behavior
becomes less pronounced with increasing pressure.  At $p$ = 12.7 kbar
between $p^{*}$ and $p_{c}$, the anomaly of $1/T_{1}$ at $T_{c}$
completely disappears, and the relaxation rate becomes smaller than
our detection limit (dotted line in Fig. 2(c)) at $p_{c} < p$.
In systems having magnetic order in a full volume fraction,
the asymmetry of the LF-$\mu$SR signal becomes 1/3 below $T_{c}$ as shown
for the results at $p$ = 8.0 kbar.  The asymmetry at $p$ = 12.7 kbar at 
$T \rightarrow 0$ is significantly larger than 1/3, indicating that the
magnetic order occurs only in a partial volume fraction, thus
confirming the results in WTF.  The full amplitude LF signal at $p$ = 16.3
kbar confirms that there is no volume of magnetically ordered region above $p_{c}$.   

The absence of any observable relaxation puts a severe
limit for the time scale of dynamic spin fluctuations of 
the ``partial order'' spin correlations.
To estimate the rate $\nu$ of this fluctuation,
here we use a well known formula, 
$1/T_{1} \sim (\gamma_{\mu}\times H)^{2}/\nu$, 
where $H$ denotes the local field strength, and
$\gamma_{\mu} = 2\pi\times 1.36 \times 10^{4}$ /sG is the 
gyromagnetic ratio of the muon spin.  
For a trial value of $H \sim$ 500 G expected for 
the Mn spin polarization of $\sim$ 0.2 $\mu_{B}$,
the lowerlimit of $\nu > 1.2 \times 10^{10}$ /s is given by
the upper limit $1/T_{1} < 0.15\mu$s$^{-1}$ indicated by the absense of 
relaxation in  WTF-$\mu$SR at $p$ = 15.0 kbar at T=2.5 K.
Similarly, the (safe) upper limit $1/T_{1} < 0.05 \mu$s$^{-1}$, 
observed in LF-$\mu$SR
at 16 kbar at T=2.9 K, indicates $\nu > 3.6 \times 10^{10}$ /s. 
Although it is difficult to obtain a precise 
value of $H$ for the partial order correlations,
the above values serve as a reasonable estimate for
the order of magnitude of $\nu$ from $\mu$SR.
Neutron scattering signals from the partial order were 
detected in a quasi-elastic 
scan with the energy resolution of
50 $\mu$eV \cite{pfleidererMnSiNature}, 
which selects static (quasi-elastic) versus dynamic responses with the
fluctuation rate of 10$^{11}$ /sec.  Thus, the combination 
of neutron and present $\mu$SR
results indicate that the ``partial order'' spin 
correlations have dynamic
character of a time scale between 10$^{-11}$ and 10$^{-10}$ s.

Since one might expect an influence of helical modulation 
in the case of MnSi, we also
performed $\mu$SR measurements in metallic 
(Sr$_{1-x}$Ca$_{x}$)RuO$_{3}$ system in ambient pressure, 
using ceramic specimens
and a surface muon beam at TRIUMF.
The unsubstituted SrRuO$_{3}$ exhibits a ferromagnetic order with 
$T_{c}\sim$ 160 K and the ordered moment $m_{s}\sim$ 0.8 $\mu_{B}$ per Ru
at $T \rightarrow 0$.  
With increasing Ca concentration $x$, both $T_{c}$ and $m_{s}$ decrease, 
and ferromagnetic order disappears
at $x\sim$ 0.7, while the paramagnetic effective 
moment $m_{p}^{eff} \sim$ 3.0 $\mu_{B}$ /Ru
remains nearly unchanged between $x$ = 0 and 1 \cite{kiyama113JPSJ}.  
This is a typical behavior expected in 
the Self Consistent Renormalization (SCR) theory of Moriya and co-workers
developed for weak ferromagnetism of itinerant 
electron systems \cite{moriyaBook}.
Figure 3(a) shows time spectra of muon spin polarization observed in zero field
at $T \sim$ 2K for specimens with various Ca concentrations.  
For $x$ = 0 and 0.5, we see
a fast damped oscillation of 2/3 of the asymmetry 
followed by a slowly relaxing 1/3
component, which is expected for systems having magnetic 
order in 100 \%\ of the volume.
The absence of long-lived muon precession is due to magnetic domain 
structures in ceramic specimens \cite{niedermayerprivate}.  The spectra 
for $x$ = 0.65 and 0.7 exhibit a slower damped oscillation,
followed by increased slow-decay component, indicating that the magnetic order
occurs in a partial volume fraction.  No clear signature of 
magnetic order can be seen for $x$ = 0.8 and 1.0 systems.

Figure 3(b) shows the local field width $\Delta$, derived from 
the damped oscillation signal, which is proportional to the ordered 
moment size within the magnetically ordered volume.
More data for around $x$ = 0.7 is needed to determine 
wheter $\Delta(T\rightarrow 0)$ changes abruptly or continuously as 
a function of $x$. 
The volume fraction $V_{f}$ of the magnetically ordered region is 
shown in Fig. 3(c), which 
clearly demonstrates phase separation between ordered and paramagnetic
regions at a narrow concentration range before disappearance of ferromagnetism.
The product of the local moment size and the volume fraction, 
shown in Fig. 3(d), scales with bulk magnetization which reflects 
the volume integrated quantity.
We have also performed $1/T_{1}$ measurements in ZF-$\mu$SR 
in (Sr$_{1-x}$Ca$_{x}$)RuO$_{3}$. The slope of $T_{1}$ 
versus $1/T$ above $T_{c}$, which signifies the strength of the critical
divergence of the relaxation rate \cite{hayanoMnSiPRL,gatMnSiPRL}, is 
reduced with increasing $x$ and becomes nearly 0 at $x\sim 0.7$, indicating 
that the dynamic critical behavior is suppressed near QPT. 
The $\mu$SR results of phase separtion
and suppression of critical dynamics in (Sr$_{1-x}$Ca$_{x}$)RuO$_{3}$ 
exhibit striking resemblance to those in MnSi,
which suggests a possibility that these features may be generic to 
QPTs in itinerant ferro/heli-magnets.

We now explore quantum crossover behaviors from magnetically 
ordered (or superconducting) to disordered states in other correlated 
electron systems.  Table 1 lists known cases where the
crossover is associated with either a first-order transition (FOT) 
or phase separation (PS) or partial volume fraction of the ordered region (PVF).  
Note that volume integrated 
measurements, such as bulk magnetization or neutron scattering, allow 
distinction of FOT only when an abrupt change is deteced at the phase 
boundary.  Continuous change
in these measurements can be due either to a true second order phase 
transition or to a FOT accompanied by PS where the product of moment size 
and volume fraction exhibits
a continuous variation as in Fig. 3(d).  In contrast, NMR/NQR and $\mu$SR can 
distinguish signals from ordered and disordered volumes.  When single crystal 
specimens are available, $\mu$SR has the further advantage of avoiding 
pressure/system heterogeneities due to powderized specimens.   
Table 1 demonstrastes that many of known cases of crossover,
not only in itinerant magnets but also in heavy-fermion and 
high-$T_{c}$ systems, are associated with FOT/PS/PVF.  It fact, it is not 
easy to find a well-established case of a QPT in such systems with 
a truly second order transition.

In his pioneering theoretical contribution which initiated modern 
discussions of QPT, Hertz \cite{hertzPRB} mapped QPT to thermal counterparts 
in a higher effective dimension,
and thus expected a better applicability of mean-field theory and second order
transitions in quantum transitions.  Contrary to this expectation,
many actual systems exhibit first order transitions.  Figure 4 illustrates 
the evolution of the free energy as a function of 
order parmeter in typical second order ((a) and (b)) 
and first order phase transitions ((c) and (d)).      
Recently, Belitz {\it et al\/} \cite{belitzPRL}
ascribed first order transition in itinerant ferromagnets to 
terms in free energy arising from coupling of low energy spin fluctuation 
modes and order parameter fluctuations, which leads to a free energy 
profile similar to that shown by the (c) line in Fig. 4 at $T_{c}$.
Alternatively, first order QPTs with such free energy profile could 
also be due, in general, to effects of band structure in metallic ferromagnets.
Randomness would suppress low energy fluctuations and/or smear out discontinuous changes, 
thus favouring tendency towards apparent second order transition in both cases. 
Clear observation of PS
in (Sr$_{1-x}$Ca$_{x}$)RuO$_{3}$, overcoming possible effects of randomness
in chemically substituted systems, can then be taken as a rather 
robust evidence against second order transition.  Although FOT and PS are 
not necessarily the identical concepts, PS is associated with FOT in 
the majority of the cases.
In general, when the relevant energy scale is lowered near a QPT, a competing 
phenomenon, such as superconductivity in UGe$_{2}$, and/or a generic
tendency towards first order transition become a dominant factor, winning 
over the simpler scheme initially proposed by Hertz.  This provides a natural 
way to understand overwhelming tendency towards FOT/PS/PVF shown in Table 1.

As a novel type of spin correlation at QPT, the partial order
correlations of MnSi have become a focus of theoretical interest, 
and several different models have been presented for their
explanation, such as 
``helical spin crystals'' \cite{binzprl},
``blue quantum fog'' \cite{tewariprl},
``skyrmion state'' \cite{roesslernature},
as well as 
``magnetic rotons'' \cite{schmalianprl}.
The magnetic roton model views the partial
order correlations as a manifestation of a spin soft mode towards
helical spin order, analogous to 
rotons in superflid He, which crosses over to 
solid helium with increasing pressure at $p \sim 26$ bar via
first-order QPT \cite{dietrichrotonPRA}.  These correlations may also be
analogous to the 41 meV neutron resonance mode in 
the cuprate systems \cite{uemurarotonJPCM,uemurarotonPhysica}
heading towards the stripe spin/charge ordered state.
By revealing the dynamic nature of the partial order
correlations, and providing their energy/time scale 10$^{10-11}$ /s, 
the present results give severe constraints to 
future development of these models / theories.

In summary, we have presented $\mu$SR studies on dynamic and static
magnetic behaviors at QPTs of MnSi and (Sr,Ca)RuO$_{3}$, which
demonstrate spontaneous phase separation and suppression of 
dynamic critical behavior.  
It is interesting to note that nearly identical spin 
responses are observed in 
MnSi which involves low crystal symmetry and Dyaloshinskii-Moriya
interaction \cite{bakjpcm} and in (Sr,Ca)RuO$_{3}$ having higher symmetry 
and ferromagnetic ground state.
These findings, together with other 
known cases in heavy-fermion and cuprate systems, promote the development of
a comprehensive understanding of QPT in correlated electron systems
which clarifies the role of first order transitions.

We acknowledge financial supports from
NSF DMR-05-02706 (Material World Network, Inter-American Materials
Collabortion programme) at Columbia and Kentucky, NSF DMR-01-02752 and CHE-01-11752 at Columbia,
NSERC and CIAR (Canada) at McMaster, Brazilian grant
CIAM-CNPq 49.2674/2004-3 at CBPF, and CIAM-CONICET proyect 509/20-04-05
at CAB Bariloche, Argentina; technical supports from S.R. Kreitzman and K. Satoh; and 
scientific discussions with B. Binz, M. Continentino, S.R. Julian, and A.J. Millis.
\\

{\bf Method\/}

In $\mu$SR \cite{scottish} in a transverse field, the time spectrum 
from a paramgnetic / nonmagnetic specimen (and pressure cell) is given by
$$N(t) = N_{o}\exp(-t/\tau_{\mu})[1+AG(t)\cos(\gamma_{\mu}H_{ext}t+\phi)],$$
where $\tau_{\mu} = 2.2 \mu$s is the muon lifetime, $\gamma_{\mu} = 2\pi\times 13.6$ kHz/G
is the gyromagnetic ratio of the muon spin, $A\sim$ 0.2-0.3 is the initial
asymmetry (constant for a given spectrometer / beamline condition), 
$G(t)$ is the relaxation function, $H_{ext}$ is the magnitude
of static external field (plus Knight shift) at the muon site, 
and $\phi$ is the initial
phase of precession (which depends on the location of the counter).
When some volume undergoes static magnetic order, producing a static internal
field $H_{int}$ at the muon site, the oscillation amplitude
for the frequency $\gamma_{\mu}H_{ext}$ is reduced.  In the case of MnSi below $T_{c}$,
where the vector sum of $H_{ext}$ (= 100 G) and $H_{int}$ ($\geq$ 900 G 
at $T \rightarrow 0$ and $p$ = 0) has a wide spread
($\sim$ 100 G or more) in magnitude, the oscillating signal from the 
magnetically ordered volume is depolarized within $t$ = 100 ns or less. 
As a display of the response from para/non-magnetic volume, Fig. 1(b) 
shows the raw data $\vert AG(t)\cos(\gamma_{\mu}H_{ext}t+\phi) \vert$,
together with the ``envelope'' $AG(t)$ obtained by dividing the data with 
the cosine function (we avoided plotting the time region where the cosine function 
is close to zero).  By fitting the amplitude of this cosine signal, assuming a slowly
decaying function for $G(t)$ (due mostly to 
nuclear dipolar fields), we obtained the fraction of muons stopped in 
the paramagnetic region of MnSi plus the non-magnetic pressure cell.  The cell contribution 
was calibrated using the data at T = 2.5 K at ambient pressure, 
where the full volume of MnSi is known to undergo static order.  Subtracting
this cell amplitude, we obtained the precession amplitude of muons representing the 
``paramagnetic volume fraction of MnSi'',
from which the plot in Fig. 2(a) was constructed.
In zero-field (ZF) or logitudinal-field (LF) $\mu$SR, the time histograms for the
forward (F) / backward (B) counters are given as
$$N_{F/B}(t) = N_{Fo/Bo}\exp(-t/\tau_{\mu})[1\pm AG(t)].$$
The relaxation function $G(t)$ describes the time evolution of muon 
spin polarization from the initial value of $G(t=0) = 1.0$.
Full relaxation makes $G(t) = 0$.  For signals from multiple different
regions, such as the cell and the specimen, $G(t)$ is decomposed into 
additive signals from different regions.  In paramagnetic systems,
the sample signal usually exhibits an exponential decay $\exp(-t/T_{1})$,
while in the ordered state in ZF one observes a damped oscillation of the 
2/3 of the asymmetry added to a slow decay of the 1/3 asymmetry,
which represents muons at the sites with the internal field
from ordered moments being rather parallel to the initial muon polarization,
as seen in $G(t)$ for $x < 0.5$ at low temperatures in Fig. 3(a).
This 2/3:1/3 ratio could depend on crystal orientation for single crystal
specimens, as well as the ratio between external and internal field
in LF.  For cubic crystal of MnSi which has eight [111] spiral directions,
however, we expect that a balanced random population of domain orientations 
at low/zero field
would minimize the orientation effect, leading to nearly 2:1 ratio
similar to the case for powder/ceramic samples. 
The asymmetry plotted in Fig. 2(c) corresponds to the 
relative magnitude of the non-oscillating
component within the signal from the MnSi specimen, after the 
background signal from the pressure cell was subtracted.

\onecolumngrid
\vfill \eject
\newpage
\begin{figure}[h]

\begin{center}
\vskip 1.0 truecm
\includegraphics[angle=0,width=7.0in]{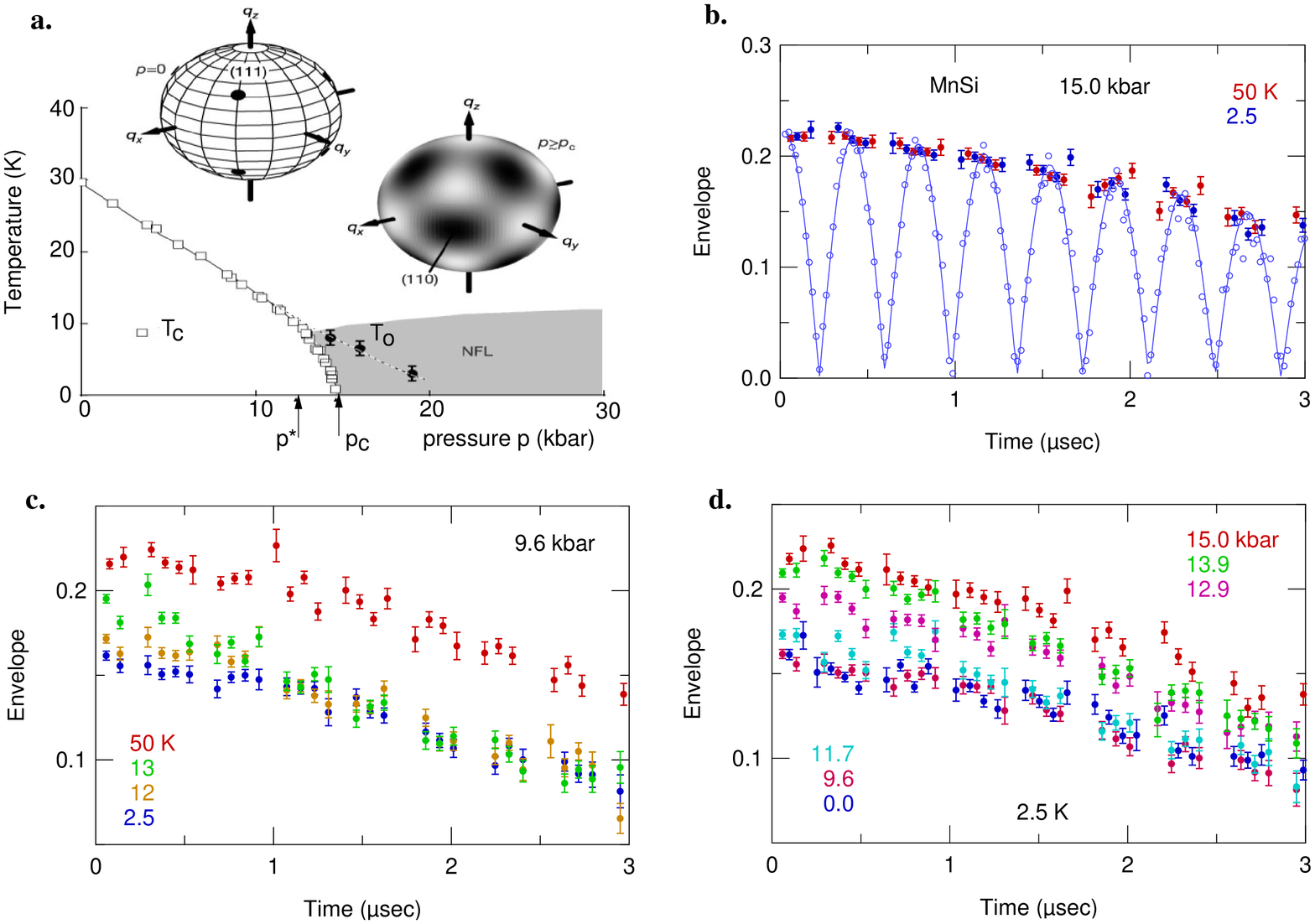}
\vskip 1.0 truecm
\label{Figure 1.} 

\caption{\label{Figure 1.}
(color)
(a) Phase diagram of MnSi as a function of pressure \cite{pfleidererMnSiNature}.
History dependent behavior has been found between $p^{*}\sim$ 12 and $p_{c}$ = 14.6 kbar.
The closed circles show a temperature $T_{o}$ below which a diffuse neutron scattering intensity,
illustrated in a right intensity-map sphere, was observed.  The shaded region exhibits
a non-Fermi-liquid behavior in transport measurements.  
(b) Muon spin precession pattern observed in a single
crystal specimen of MnSi within a pressure cell in a weak transverse external field 
(WTF) of 100 G, and the envelope of the oscillation spectra at T = 50 K and 2.5 K at 
$p$ = 15 kbar.  The amplitude of this envelope represents muons in the non-magnetic or
paramagnetic environment.  No difference between T = 50 K and 
2.5 K indicates absence of static magnetic order at T = 2.5 K at a pressure slightly above $p_{c}$.
(c) Temperature dependence of the envelope at $p$ = 9.6 kbar.  The reduction of the precessing
envelope is caused by the volume of the specimen which has undergone static magnetic order.
The envelope observed at T = 2.5 K represents muons stopped in 
the pressure cell, after the signal from ordered MnSi being quickly 
depolarized.  
(d) Pressure dependence of the envelope at T = 2.5 K.  At ambient pressure and 9.6 kbar,
100 \%\ of the volume of MnSi undergoes static magnetic order, while the static magnetic
order takes place in a reduced volume fraction between $p$ = 11.7 and 13.9 kbar.} 
\end{center}
\end{figure}
\vfill \eject
\newpage

\begin{figure}[t]

\begin{center}

\includegraphics[angle=0,width=6.5in]{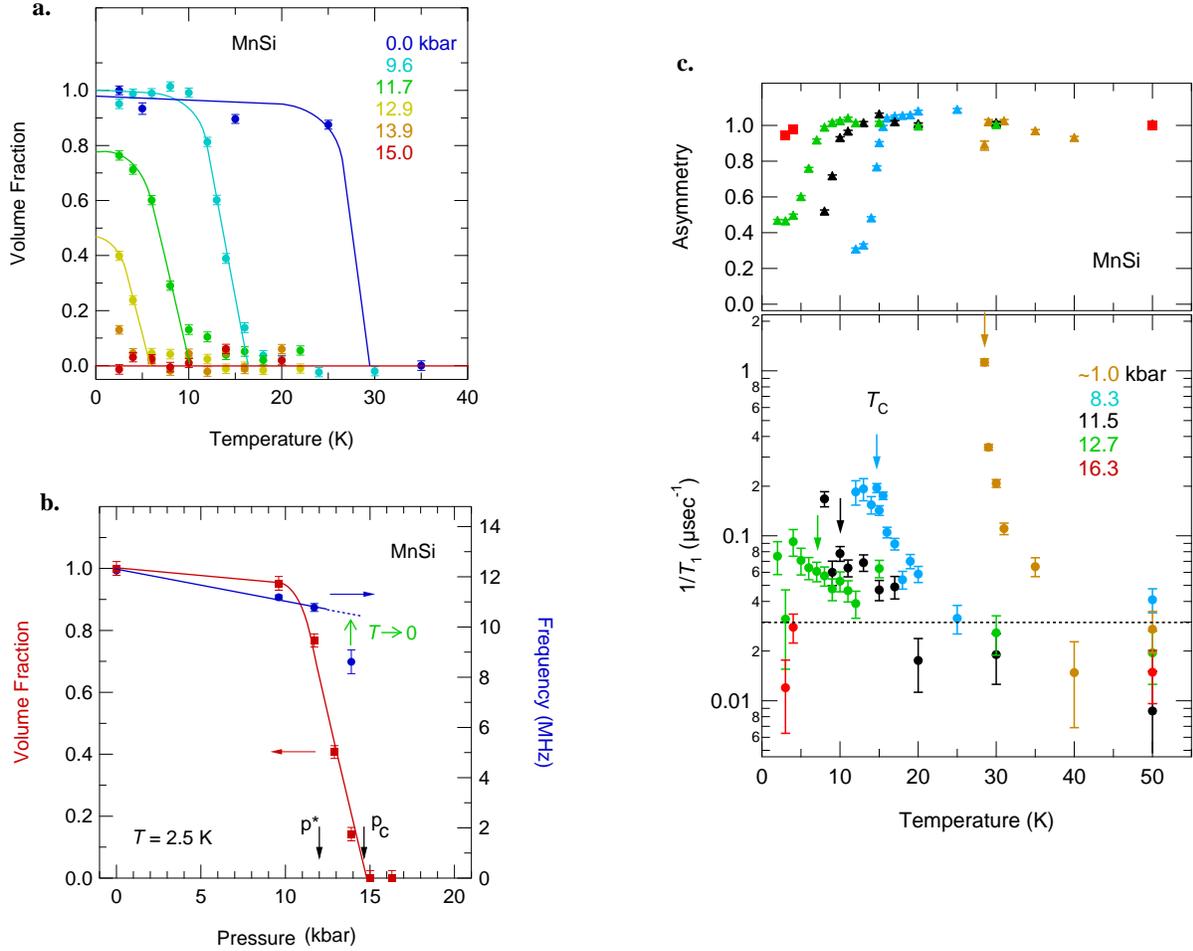}
\vskip 1.0 truecm 

\label{Figure 2.} 

\caption{\label{Figure 2.} 
(color) 
(a) Temperature and pressure dependences of the volume fraction $V_{M}$ having
static magnetic order in MnSi determined from the muon precession envelope
measured in WTF of 100 G.  The volume fraction remains finite at $T \rightarrow 0$
at the pressure $p$ between 11.7 and 13.9 kbar, indicating phase separation 
between magnetically ordered and paramagnetic volumes.
(b) Pressure dependence of the ordered volume fraction, determined
in WTF of 100 G, and the muon spin precession frequency observed in zero-field 
$\mu$SR, at T = 2.5 K.  The finite frequency near $p_{c}$ indicates 
a first order phase transition.  The frequency at $p$ = 13.9 kbar at T = 2.5 K $\sim$
0.5 $T_{c}$ is expected
to increase at $T \rightarrow 0$ as illustrated by the green arrow.
(c) The muon spin relaxation rate $1/T_{1}$ and the relaxing muon asymmetry
obtained in $\mu$SR measurements in a longitudinal field (LF) of 200 G.  
Divergent critical behavior of $1/T_{1}$, seen at $p\sim$ 1 kbar, is 
gradually suppressed with increasing pressure.  No anomaly of $1/T_{1}$ is 
seen at $T_{c}$ (indicated by arrows) at $p$ = 12.7 kbar, between $p^{*}$
and $p_{c}$. At $p$ = 16.3 kbar, $1/T_{1}$ becomes smaller than the 
technical limit of detection, indicated by the broken line.
} 
\end{center}
\end{figure}

\newpage

\begin{figure}[t]

\begin{center}

\includegraphics[angle=0,width=7.0in]{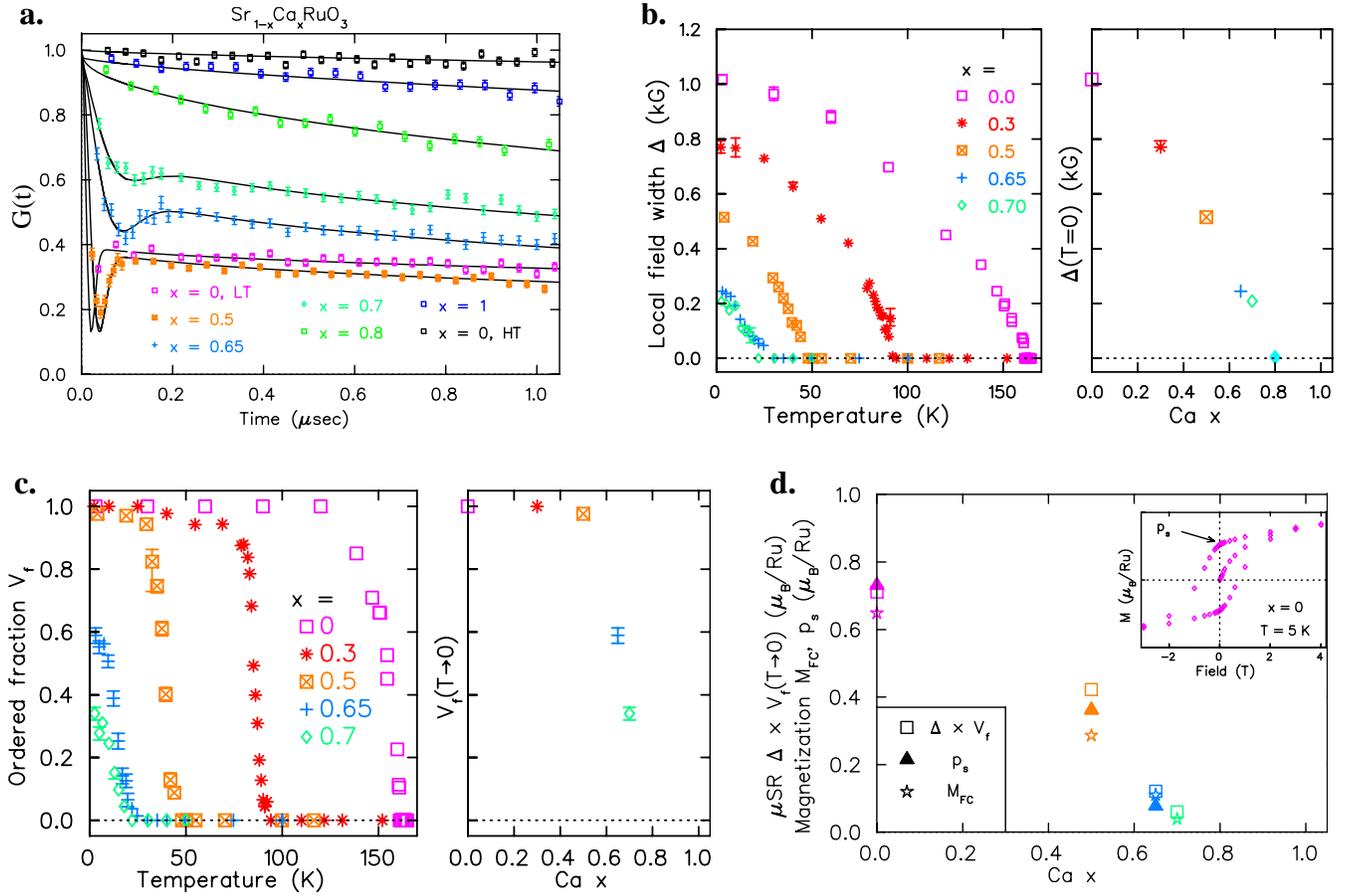}
\vskip 1.0 truecm

\label{Figure 3.} 

\caption{\label{Figure 3.}
(color)
(a) Muon spin relaxation spectra observed in zero field in (Sr$_{1-x}$Ca$_{x}$)RuO$_{3}$
at T $\sim$ 2.5 - 5 K (indicated as LT) in ceramic specimens of various $x$,
compared with that at T = 164 K (HT) in SrRuO$_{3}$ above $T_{c}$.  The LT spectra
shows a damped oscillation with the amplitude of 2/3, expected for
static order in full volume, in $x$ = 1.0 and 0.5. 
Reduced amplitue is seen for $x$ = 0.65 and 0.7, indicating a finite volume 
fraction of magnetically ordered region. The oscillation is replaced by a slow 
relaxation in $x$ = 0.8 - 1.0.  The solid line represents a fit to a Gaussian
Kubo-Toyabe function multiplied by an exponential decay.
(b) Temperature and concentration dependence of the amplitude $\Delta$ of static
random local field, derived from the fit of ZF-$\mu$SR data to 
the Gaussian Kubo Toyabe function, in (Sr$_{1-x}$Ca$_{x}$)RuO$_{3}$.
The width $\Delta$ is proportional to the average size of static ordered moment in the 
magnetically ordered volume.  Note that $\Delta (T\rightarrow 0)$ shows a marked reduction from 
$x$ = 0.7 to 0.8.
(c) Volume fraction $V_{f}$ of the magnetically ordered region determined by 
ZF-$\mu$SR in (Sr$_{1-x}$Ca$_{x}$)RuO$_{3}$.  For $x >$ 0.7, the damped oscillation 
is replaced by a slow relaxation, which suggests disappearence of the volume 
having ordered moment greater than $m_{s} \sim$ 0.01 $\mu_{B}$ per Ru.
(d) Comparison of $\Delta \times V_{f}(T\rightarrow 0)$ by $\mu$SR to 
magnetization measured in field cooling ($M_{FC}$) and the spontaeous moment $p_{s}$
seen in the field-cycling, as illustrated in the inset,
all measured at low temperatures (2-5) K.  The latter two 
quantities represent the volume integrated response.  Reasonable agreement of 
those with the product of $\Delta$ and $V_{f}$ further confirms phase separation
before static magnetism disappears around $x \sim$ 0.7.} 

\end{center}
\end{figure}
\newpage
\vfill \eject
\newpage

\begin{figure}[t]

\begin{center}

\includegraphics[angle=0,width=5.5in]{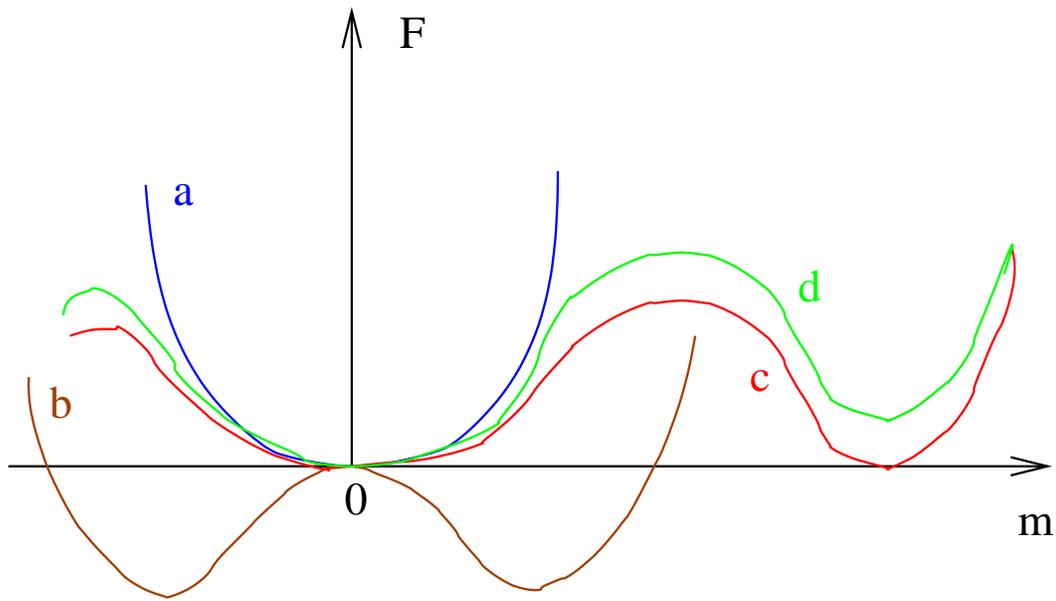}
\vskip 1.0 truecm 

\label{Figure 4.} 

\caption{\label{Figure 4.} 
(color) 
A schematic view of free energy $F$, as a function of magnetic
order parameter $m$, illustrated for the second order
phase transition in the disordered state (line (a)); and the ordered state
(line (b)); and the first order phase transition at the ordering
temperature or pressure (line ((c)) and in the disordered state near
the ordering (line(d)).  A discussion for the situation with line (c) in 
itinerant ferromagnets has been given by Belitz {\it et al.\/}
\cite{belitzPRL}.}    

\end{center}
\end{figure}
\vfill \eject
\newpage
\squeezetable
\begin{table*}[t] 
 \caption{\label{datatable}
Phase Separation  (PS) / First-Order Transition (FOT) / Partial-Volume (PVF) features detected at QPT}
 \begin{ruledtabular}
 \begin{tabular}{llcccccc}
System & Crossover & Parameter & FOT/PS/PVF & Method & Ref. & Remark, \ [Limitation]&\\
       &           & ($p$ in kbar)&       &        &      & \\         
\hline
   \\
{\sl Itinerant electron magnets\/}\\
   \\
UGe$_{2}$&FM1-Para&$p_{c}$=16    &FOT     &Magnetization&\cite{pfleidererhuxley}&Sudden drop&  \\
         &        &              &PS     &Ge-NQR&\cite{haradanqr}&coexist signal\ [powder]\\
   \\
ZrZn$_{2}$  &Ferro-Para&$p_{c}$ =16.5 &FOT     &Magnetization&\cite{uhlarzZrZn2PRL}&Sudden drop& \\
   \\
MnSi&Helical-Para&12=$p^{*}<p<p_{c}$=14.6&FOT&Susceptibility&\cite{pfleidererMnSiPRB}&sharp change& \\
    &            & $p<p_{c}$       &FOT& Si-NMR&\cite{thessieuMnSiPhysica}& [powder] & \\      
    &            & $p<18$       &PVF     &Si-NMR&\cite{yuMnSiPRL}&intensity drop\ [powder] \\
    &            & $p^{*}<p<p_{c}$&PS&$\mu$SR&Present Work&volume fraction,\ x-tal\\
    &            &                &FOT&       &            &suppressed critical dynamics\\
   \\
(Sr$_{1-x}$Ca$_{x}$)RuO$_{3}$&Ferro-Para&0.7$<$Ca($x$)$<$0.65&PS&$\mu$SR&Present Work
&volume fraction\\
    &            &                &FOT&       &            &suppressed critical dynamics\\
   \\
\hline
   \\
{\sl Heavy Fermion Systems\/}\\
   \\   
CeCu$_{2.2}$Si$_{2}$&AF(a)-SC& Temp. &PS&$\mu$SR&\cite{lukeCeCu2Si2PRL}&AF volume change at SC& \\
   \\
URu$_{2}$Si$_{2}$&AF-Hidden&3$<p<$10&PVF&Si-NMR&\cite{matsudaURu2Si2PRL}& [hidden order missing]\\
  & &0$<p<$10&PVF/FOT&$\mu$SR&\cite{lukeURu2Si2HFI,amitsukaURu2Si2Physica}&\\
   \\
CeIn$_{3}$&AF-SC&$p_{c}$=24.3&FOT&In-NQR&\cite{kitaokaHFJPSJ}&FOT also in FM-PM\\
   \\
\hline
   \\
{\sl High-T$_{c}$ systems\/}\\
   \\
(La,Sr)$_{2}$(Cu$_{1-x}$Zn$_{x}$)O$_{4}$&SC-normal& Zn($x$)&PS&$\mu$SR&\cite{nachumiHTSCZnPRL}&Swiss Cheese model&\\
Bi$_{2}$Sr$_{2}$Ca(Cu$_{2-x}$Zn$_{x}$)O$_{8}$&SC-normal& Zn($x)$&PS&STM&\cite{panHTSCZnNature}&normal 
region around Zn\\
   \\
(La$_{1.85-y}$Sr$_{0.15}$Eu$_{y}$)CuO$_{4}$&Stripe-SC& Eu($y$) &PS&$\mu$SR&\cite{kojimaLESCOPhysica}&
 volume fraction trade-off\\
   \\
O-doped (La$_{2-x}$Sr$_{x}$)CuO$_{4+y}$&Stripe-SC&Temp.,\ Sr($x$)&PVF/PS&$\mu$SR
&\cite{saviciLCOPRB,mohottalaLSCONaturePhys}&stripe islands\cite{saviciLCOPRB}\\
   \\
Tl$_{2}$Ba$_{2}$CuO$_{6+\delta}$&SC-Metal&overdope O($\delta$)&PS&$\mu$SR&
\cite{uemuraTl2201Nature,uemurarotonJPCM}&boson-fermion coexistence\\ 
Bi$_{2}$Sr$_{2}$CaCu$_{2}$O$_{8+\delta}$&SC-Metal&overdope O($\delta$)&PS&STM&\cite{davisprivate}&inhomogeneous gap closing\\
\\
 \end{tabular}
 \end{ruledtabular}
\\
\ \ \\
SC denotes superconducting phase, STM denotes Scanning Tunnelling Microscope.
 \end{table*}

\vfill \eject
\newpage

\begin{thebibliography}{99}

\bibitem{pfleidererMnSiPRB}
C. Pfleiderer, G. J. McMullan, S. R. Julian, and G. G. Lonzarich,
{Magnetic quantum phase transition in MnSi under hydrostatic pressure},
Phys. Rev. {\bf B 55\/} (1997) 8330 - 8338.
\bibitem{pfleidererMnSiNature}
C. Pfleiderer, D. Reznik, L. Pintschovius, H.v. L\"ohneysen,
M. Garst, A. Rosch,
{Partial order in the non-Fermi-liquid phase of MnSi},
Nature {\bf 427\/} (2004) 227 - 231.
\bibitem{pfleidererZrZn2Nature}
C. Pfieiderer,  M Uhlarz,  S M Hayden,  R Vollmer, H. v. L\"ohneysen, N. R. Bernhoeft, 
G. G. Lonzarich,
{Coexistence of superconductivity and ferromagnetism in the d-band metal ZrZn$_{2}$},
Nature {\bf 412\/} (2001) 58-61.
\bibitem{saxenaUGe2Nature}
S.S. Saxena, P. Agarwal, K. Ahilan, F.M. Grosche, R.K.W. Haselwimmer, M.J. Steiner, 
E. Pugh, I.R. Walker, S.R. Julian, P. Monthoux, 
G.G. Lonzarich, A. Huxley, I. Sheikin, D. Braithwaite, J. Flouque, 
{Superconductivity on the border of itinerant-electron ferromagnetism
in UGe$_{2}$}, Nature {\bf 406\/} (2000) 587-592.
\bibitem{ishikawaMnSiPRB}
Y. Ishikawa, G. Shirane, J. A. Tarvin, and M. Kohgi, 
{Magnetic excitations in the weak itinerant ferromagnet MnSi}, 
Phys. Rev. {\bf B16\/} (1977) 4956-4970.
\bibitem{moriyaBook}
T. Moriya, {\bf Spin Fluctuations in Itinerant Electron Magnetism\/}, Springer
Series in Solid State Science {\bf 56\/}, (Springer, Heidelberg, 1985),
and references therein.
\bibitem{thessieuMnSiPhysica}
C. Thessieu, Y. Kitaoka and K. Asayama,
{Magnetic quantum phase transition in MnSi\/},
Physica {\bf B259-261} (1999) 847-848. 
\bibitem{yuMnSiPRL}
W. Yu, F. Zamborszky, J.D. Thompson, J.L. Sarrao,
M.E. Torelli, Z. Fisk, and S.E. Brown,
{Phase Inhomogeneity of the Itinerant Ferromagnet MnSi at High Pressures},
Phys. Rev. Lett. {\bf 92\/} (2004) 086403 [4 pages].
\bibitem{scottish} For a recent reviews of $\mu$SR studies in topical
subjects and technical aspects of $\mu$SR, 
see {\it Muon Science: Muons in Physics, Chemistry and
Materials\/}, Proceedings of the Fifty First Scottish 
Universities Summer School in Physics, St. Andrews, August, 1988, 
ed. by S.L. Lee, S.H. Kilcoyne, and R. Cywinski,
Inst. of Physics Publishing, Bristol, 1999.
\bibitem{saviciLCOPRB}
A.T. Savici, Y. Fudamoto, I.M. Gat, T. Ito, M.I. Larkin, Y.J. Uemura,
G.M. Luke, K.M. Kojima, Y.S. Lee,  M.A. Kastner,
R.J. Birgeneau, K. Yamada, 
{Muon spin relaxation studies of incommensurate magnetism and 
superconductivity in stage-4
La$_{2}$CuO$_{4.11}$ and La$_{1.88}$Sr$_{0.12}$CuO$_{4}$\/}, 
Phys. Rev. {\bf B66\/} (2002) 014524. 
\bibitem{hayanoMnSiPRL}
R.S. Hayano, Y.J. Uemura, J. Imazato, N. Nishida, 
T. Yamazaki, H. Yasuoka, and Y. Ishikawa,
{Observation of the $T / (T-T_{c})$ Divergence of the $\mu^{+}$ 
Spin-Lattice Relaxation Rate in MnSi near $T_{c}$},
Phys. Rev. Lett. {\bf 41\/} (1978) 1743 - 1746.
\bibitem{kadonoMnSiPRB}
R. Kadono, T. Matsuzaki, T. Yamazaki, S.R. Kreitzman, and J.H. Brewer,
{Spin dynamics of the itinerant helimagnet MnSi studied by positive muon spin relaxation},
Phys. Rev. {\bf B42\/} (1990), 6515-6522.
\bibitem{gatMnSiPRL}
I.M. Gat-Malureanu, A. Fukaya, M.I. Larkin, A.J. Millis, P.L. Russo, A.T. Savici, 
Y.J. Uemura, P.P. Kyriakou, G.M. Luke, C.R. Wiebe, Y.V. Sushko, R.H. Heffner, 
D.E. MacLaughlin, D. Andreica, and G.M. Kalvius,
{Field Dependence of the Muon Spin Relaxation Rate in MnSi},
Phys. Rev. Lett. {\bf 90\/} (2003) 157201 [4 pages].
\bibitem{kiyama113JPSJ}
T. Kiyama, K. Yoshimura, K. Kosuge, H. Mitamura, T. Goto,
{High-Field Magnetization of Sr$_{1-x}$Ca$_{x}$RuO$_{3}$},
J. Phys. Soc. Japan {\bf 68\/} (1999) 3372-3376.
\bibitem{niedermayerprivate}
Ch. Niedermayer, private communication, 2006.
\bibitem{hertzPRB}
J.A. Hertz, {Quantum critical phenomena},
Phys. Rev. {\bf B14\/} (1976) 1165-1184.  
\bibitem{belitzPRL}
D. Belitz, T. R. Kirkpatrick, and J. Rollb\"uhler,
{Tricritical Behavior in Itinerant Quantum Ferromagnets},
Phys. Rev. Lett. {\bf 94\/} (2005) 247205.
\bibitem{binzprl}
B. Binz, A. Vishwanath, and V. Aji, 
{Theory of the helical spin crystal: A candidate for the partially ordered 
state of MnSi}, Phys. Rev. Lett, {\bf 96\/} (2006) 207202.
\bibitem{tewariprl}
S. Tewari, D. Belitz and T. R. Kirkpatrick,
{Blue quantum fog: Chiral condensation in quantum helimagnets},
Phys. Rev. Lett. {\bf 96\/} (2006) 047207.
\bibitem{roesslernature}
U. K. R\"ossler1, A. N. Bogdanov, and C. Pfleiderer,
{Spontaneous skyrmion ground states in magnetic metals},
Nature (2006), in press.
\bibitem{schmalianprl}
J. Schmalian and M. Turlakov, {Quantum phase transitions of magnetic rotons},
Phys. Rev. Lett. {\bf 93\/} (2004) 036405.
\bibitem{dietrichrotonPRA}
O.W. Dietrich, E.H. Graf, C.H. Huang,
L. Passell,
{\sl Neutron scattering by rotons in 
liquid helium\/},
Phys. Rev. {\bf A5\/} (1972) 1377.
\bibitem{uemurarotonJPCM}
Y.J. Uemura,
{Condensation, excitation, pairing, and superfluid density in high-$T_{c}$ 
superconductors:
magnetic resonance mode as a roton analogue and 
a possible spin-mediated pairing},
J. Phys. Condens. Matter {\bf 16\/} (2004) S4515 - S4540.
\bibitem{uemurarotonPhysica}
Y.J. Uemura, {Twin spin/charge roton mode and superfluid density: Primary
determining factors of Tc in high-Tc superconductors observed by neutron, ARPES, and MuSR},
Physica {\bf B374-375\/} (2006) 1-8.
\bibitem{bakjpcm}
P. Bak and M. H. Jensen, 
{Theory of helical magnetic structures and phase
transitions in MnSi and FeGe\/},
J. Phys. C: Solid St. Phys. {\bf 13\/} (1980) L881-5.
\bibitem{pfleidererhuxley}
C. Pfleiderer and A.D. Huxley,
{Pressure dependence of the magnetization in the ferromagnetic superconductor UGe$_{2}$},
Phys. Rev. Lett. {\bf 89\/} (2002) 147005.  
\bibitem{haradanqr}
A. Harada, S. Kawasaki, H. Kotegawa, Y. Kitaoka, Y. Haga, E. Yamamoto, Y. Onuki, 
K.M. Itoh, E.E. Haller, and H. Harima,
{Cooperative Phenomenon of Ferromagnetism and Unconventional Superconductivity
in UGe$_{2}$: A $^{73}$Ge-NQR Study under Pressure},
J. Phys. Soc. Jpn. {\bf 74\/} (2005) 2675.
\bibitem{uhlarzZrZn2PRL}
M. Uhlarz, C. Pfleiderer, and S. M. Hayden,
{Quantum Phase Transitions in the Itinerant Ferromagnet ZrZn$_{2}$},
Phys. Rev. Lett. {\bf 93\/} (2004) 256404.
\bibitem{lukeCeCu2Si2PRL}
G.M. Luke, A. Keren, K. Kojima, L.P. Le, B.J. Sternlieb, 
W.D. Wu, Y.J. Uemura, Y. Onuki and T. Komatsubara,
{Competition between Magnetic Order and Superconductivity in CeCu$_{2.2}$Si$_{2}$}, 
Phys. Rev. Lett. {\bf 73\/} (1994) 1853-1856
\bibitem{matsudaURu2Si2PRL}
K. Matsuda, Y. Kohori, T. Kohara, K. Kuwahara, and H. Amitsuka,
{Spatially Inhomogeneous Development of Antiferromagnetism in 
URu$_{2}$Si$_{2}$: Evidence from $^{29}$Si NMR under Pressure},
Phys. Rev. Lett. {\bf 87\/} (2001) 087203 (2001)
\bibitem{lukeURu2Si2HFI}
G.M. Luke, A. Keren, L.P. Le, W.D. Wu, Y.J. Uemura,
D. Bonn, L. Taillefer, J.D. Garrett, Y. Onuki,
{Muon Spin Relaxation in Heavy Fermion Systems},
Hyperfine Interact. {\bf 85\/} (1994) 397-409.
\bibitem{amitsukaURu2Si2Physica}
H. Amitsuka, M. Yokoyama, S. Miyazaki, K. Tenya, T. Sakakibara, 
W. Higemoto, K. Nagamine, K. Matsuda, Y. Kohori and T. Kohara, 
{Hidden order and weak antiferromagnetism in URu$_{2}$Si$_{2}$},
Physica {\bf B 312\/}-{\bf 313\/} (2002) 390-396.
\bibitem{kitaokaHFJPSJ}
Y. Kitaoka, S. Kawasaki, T. Mito, Y. KAwasaki,
{Unconventional Superconductivity in Heavy Fermion Systems},
J. Phys. Soc. Japan {\bf 74\/} (2005) 186, and reference therein.
\bibitem{nachumiHTSCZnPRL}
B. Nachumi, A. Keren, K. Kojima, M. Larkin, G.M. Luke, J. Merrin,
O. Tchernyshov, Y.J. Uemura, N. Ichikawa, M. Goto, and S. Uchida, 
{Muon Spin Relaxation Studies of Zn-Substitution Effects
in High-$T_{c}$ Cuprate Superconductors}, 
Phys. Rev. Lett. {\bf 77\/} (1996) 5421-5424. 
\bibitem{panHTSCZnNature}
S.H. Pan, E.W. Hudson, K.M. Lang, H. Eisaki, S. Uchida, J.C. Davis,
{Imaging the Effects of Individual Zinc
Impurity Atoms on Superconductivity in 
Bi$_{2}$Sr$_{2}$CaCu$_{2}$O$_{8+\delta}$}, 
Nature (London) {\bf 403\/} (2000) 746.
\bibitem{kojimaLESCOPhysica}
K.M. Kojima, S. Uchida, Y. Fudamoto, I.M. Gat, M.I. Larkin,
Y.J. Uemura, G.M. Luke,
{Superfluid density and volume fraction of static 
magnetism in stripe-stabilized La$_{1.85-y}$Cu$_{y}$Sr$_{0.15}$CuO$_{4}$},
Physica {\bf B 326\/} (2003) 316-320.
\bibitem{mohottalaLSCONaturePhys}
H.E. Mohottala, B.O. Wells, J.I. Budnick, W.A. Hines,
Ch. Niedermayer, L. Udby, C. Bernhardt, A.R. Moodenbaugh, 
F.-C. Chou,
{Phase separation in superoxygenated
La$_{2-x}$Sr$_{x}$CuO$_{4+y}$},
Nature Materials {\bf 5\/} (2006) 377-382.
\bibitem{uemuraTl2201Nature}
Y.J. Uemura, A. Keren, L.P. Le, G.M. Luke, W.D. Wu, Y. Kubo, T. Manako,
Y. Shimakawa, M. Subramanian, J.L. Cobb, and J.T. Markert,
{Magnetic Field Penetration Depth in 
Tl$_{2}$Ba$_{2}$CuO$_{6+\delta}$ in the Overdoped Regime},
Nature {\bf 364\/} (1993) 605-607.
\bibitem{davisprivate}
J.C. Seamus Davis, private communication (2005).
\end{thebibliography}
\end{document}